\documentclass[aps,prl,twocolumn,showpacs,amsmath,amssymb,floatfix]{revtex4}
\usepackage{graphicx}
\usepackage{bm}

 \draft

\begin{document}
\title{Thermal Lifshitz Force between Atom and Conductor with Small Density of Carriers}
\author{L.~P.~Pitaevskii}
\date{\today}
\affiliation{CNR INFM-BEC and Departament of Physics, University of Trento,
I-38050 Povo, Trento, Italy; \\
Kapitza Institute for Physical Problems, ul. Kosygina 2, 119334
Moscow, Russia}

\begin{abstract}A new theory describing the interaction between atoms and a conductor with small densities of current carriers is presented. The theory takes into account the penetration of the static component of the thermally fluctuating field in the conductor and generalizes the Lifshitz theory in the presence of a spatial dispersion. The equation obtained for the force describes the continuous crossover between the Lifshitz results for dielectrics and metals.
\end{abstract}
\maketitle

{\it Introduction.} The forces acting on atoms near the surface of a dielectric body was calculated at arbitrary distances and temperatures
by E.~Lifshitz  \cite{Lif}. The Lifshitz theory is based on the calculation of the stress tensor for fluctuations of the electromagnetic field between two bodies. This tensor can be expressed in terms of the complex dielectric permittivities of the bodies at imaginary values of frequency $\omega $. The general equation for atom-surface forces was obtained as a limiting case of interaction between a dielectric body 1 with dielectric permittivity $%
\varepsilon \left( \omega \right) $ and a body 2, in the limit when the second body is considered dilute from the electrodynamic point of view, i.e. assuming that dielectric function of the second body $\varepsilon_{2}\left( \omega \right) \approx 1$ and expanding with respect to $\varepsilon _{2}-1$. In general, both zero-point and thermal fluctuations contribute to the force. In this letter we will discuss only the situation when the main contribution is due to the thermal fluctuations, as it occurs at hight enough temperature and large distance $l$ between the atom and surface of the body.

It was discovered by Lifshitz, that under the condition
\begin{equation}
l\gg \lambda _{k_BT}\equiv \hbar c/(k_BT)  \label{Lif} \ ,
\end{equation}
the interaction energy decreases according to the $1/l^{3}$ law:
\begin{equation}
V\left( l\right) =-\frac{k_BT}{4l^{3}}\alpha \left( 0\right) \frac{%
\varepsilon \left( 0\right) -1}{\varepsilon \left( 0\right)+1 }\ ,
\label{VLif}
\end{equation}
where $\alpha \left( 0 \right) $ and $\varepsilon \left( 0 \right)$ are, correspondingly, the static values of the electric polarizability of an atom and the dielectric permittivity of the body.

This simple equation (\ref{VLif}), however, exhibits peculiar properties when applied to a conductor. Electrodynamic properties of a conductor at low frequencies can be described by a universal equation for the complex dielectric permittivity:
\begin{equation}
\varepsilon \left( \omega \right) =i\frac{4 \pi \sigma}{\omega}+{\bar \varepsilon} \ ,
\label{Drude}
\end{equation}
where $\sigma $ is the d.c. conductivity and ${\bar \varepsilon}$ is the ``bare'' dielectric constant, which does not take into account the contribution from current carriers. Because $\varepsilon \left( \omega \right)$ from (\ref{Drude}) tends to infinity as $\omega \to 0$ for any value of $\sigma $, the potential tends to the universal limit
\begin{equation}
V\left( l\right) =-\frac{k_BT}{4l^{3}}\alpha \left( 0\right) \ ,
\label{VLifcond}
\end{equation}
which does not depend on ${\bar \varepsilon}$ for an arbitrarily small conductivity - a seeming contradiction to common sense. However, the universality of (\ref{VLifcond}) has a simple physical meaning. It expresses the fact that the static electric field does not penetrate into any conductor. Still, this statement, accepted in macroscopic electrodynamics, is only an approximate one. Actually, the field penetrates into conductors to a depth on the order of the well-known Debye radius $R_{D}$. For good conductors it typically is of the order of interatomic distances. However, if the density of current carriers in the conductor is small, $R_{D}$ can be large. We will see below that (\ref{VLifcond}) is valid only at the condition
\begin{equation}
R_{D} \ll l \ .
\label{Main}
\end{equation}
(The problem of the temperature dependence of the forces is a subject of active discussion. See \cite{Most,Brevik} and references therein.)


{\it Atom-surface interaction.} As a first step we will derive the Lifshitz expression for the atom-surface interaction directly from the Matsubara Green's function formalism of \cite{DP} (see also \cite{LP9}).
The basis of this theory is the equation for the variation $\delta F$ of the free energy for a small change of the dielectric
permittivity $\delta \varepsilon :$%
\begin{equation}
\delta F=\frac{k_BT}{4\pi \hbar }\sum\limits_{s=0}^{\infty }\;'\int {\cal D}%
_{ll}^{E}\left( \zeta _{s};{\bf r},{\bf r}\right) \delta \varepsilon
(i\left| \zeta _{s}\right| ,{\bf r})d^{3}x  \ ,\label{dF}
\end{equation}
where ${\cal D}_{lk}^{E}$ is the Matsubara Green's function of the electric field, $\zeta _{s}=2s\pi k_BT/\hbar $ and sign `` $'$ '' means that the $s=0$ term is taken with the coefficient $1/2.$ Notice now, that the presence of an atom at the point ${\bf r}_{a}$ can be considered as a small change of the dielectric permittivity
\begin{equation}
\delta \varepsilon (\omega ,{\bf r})=4\pi \alpha \left( \omega \right)
\delta \left( {\bf r}-{\bf r}_{a}\right) .  \label{de}
\end{equation}
Substitution of (\ref{de}) into (\ref{dF}) gives the final expression for the energy of the atom-surface interaction
\begin{equation}
V(l)=\frac{k_BT}{\hbar}\sum\limits_{s=0}^{\infty }\alpha (i\left| \zeta _{s}\right| )\left[
{\cal D}_{ll}^{E}\left( \zeta _{s};{\bf r},{\bf r}^{\prime }\right) \right]
_{{\bf r}\rightarrow {\bf r}^{\prime }\rightarrow {\bf r}_{a}}\ ,  \label{Vgen}
\end{equation}
where ${\cal D}_{ik}^{E}\left( \zeta _{s};{\bf r},{\bf r}^{\prime }\right) $ is the Matsubara Green's function of the electric field for a dielectric half-space. 
(The interaction can be analogously expressed in terms of the real-frequencies Green's functions, see \cite{real}.)
Stress that (\ref{Vgen}) is an exact equation. The problem is to calculate ${\cal D}^{E}_{ik}$ in a proper approximation, in our case taking into account spatial dispersion of dielectric.
The function ${\cal D}_{ik}^{E}$ satisfies the equation
\begin{equation}
\begin{split}
\left[ \partial _{i}\partial _{l}-\delta _{il}\Delta +\frac{\zeta _{s}^{2}}{%
c^{2}}\varepsilon (i\left| \zeta _{s}\right|, {\bf r})\delta _{il}\right]
{\cal D}_{lk}^{E}\\
 =4\pi
\hbar \frac{\zeta _{s}^{2}}{c^{2}}\delta _{ik}\delta \left( {\bf r}-{\bf r}%
^{\prime }\right) .  
\label{Eq}
\end{split}
\end{equation}
In media with spatial dispersion, which will be considered below, $\varepsilon (i\left| \zeta _{s}\right| ,{\bf r})$ must be understood as a linear operator, actig on the variable ${\bf r}$.
Equation (\ref{Vgen}) allows one to obtain the general Lifshitz result for the atom-surface interaction in a much more direct way than in the original Lifshitz paper because it is much easier to solve (\ref{Eq}) for a half-space than for the two-body geometry of \cite{Lif}. We will see that by using (\ref{Vgen}) one can obtain the limiting equation (\ref{VLif}) in a very simple way. The point is that relativistic retardation effects are not important under conditions (\ref{Lif}).  Indeed, the general condition of neglecting retardation, i.e. the condition of the quasi-stationarity of the field, is
\begin{equation}
\omega \ll c/l.  \label{quasi}
\end{equation}
At the ``Lifshitz distances'' (\ref{Lif}) only the $s=0$ term in (\ref{Vgen}) is important. The problem then becomes purely static and one may calculate the Green's function, neglecting relativistic retardation effects. This problem will be solved in the next section.  Notice that the above-mentioned peculiarity of the small $\sigma$ limit exists only for the $s=0$ term.

{\it Longitudinal Green's function.} As a first step we will
separate the Green's function into its longitudinal and transverse parts, i.e.  will present it in the form
\begin{equation}
{\cal D}_{ik}^{E}={\cal D}_{ik}^{EL}+{\cal D}_{ik}^{ET}
\end{equation}
where
\begin{equation}
\left( {\rm rot}\right) _{il}{\cal D}_{lk}^{EL}=0\ {\rm and\ \partial }_{i}%
{\cal D}_{ik}^{ET}=0\ .  \label{LT}
\end{equation}

The central point of the derivation is the statement that the smallness of retarding effects implies the inequality
\begin{equation}
\left| {\cal D}_{ik}^{ET}\right| \ll \left| {\cal D}_{ik}^{EL}\right|\ .
\label{DlDt}
\end{equation}
This means that only fluctuations of electrostatic nature are important. Indeed, a simple estimate from (\ref{Eq}) gives
${\cal D}^{ET}\sim \left( l\zeta _{s}/c\right) ^{2}{\cal D}^{EL}\ll {\cal D}^{EL}.$ Particularly, the leading $s=0$ term in
(\ref{Vgen}) is defined only by 
 the longitudinal function ${\cal D}_{ik}^{EL}\left( \zeta _{s};{\bf r,r}^{\prime }\right) $.

To obtain an equation for the longitudinal part, let us apply the operator $\partial _{i}$ to both sides of $\left( \ref{Eq}\right) $. Neglecting the term of the order of $\zeta^2_s {\cal D}^{ET}$, we find
\begin{equation}
\partial _{i}\left [\varepsilon D_{ik}^{EL}\right]
=-4\pi \hbar \partial _{k}^{\prime }\delta \left( {\bf r-r}^{\prime }\right)\ .
\label{EqD1}
\end{equation}
The first equation (\ref{LT}) can be satisfied identically if we introduce a scalar function $\varphi $ according to
\begin{equation}
D_{ik}^{E}\approx D_{ik}^{EL }=\hbar \partial _{i}\partial_{k}^{\prime }\varphi \ .  \label{phi}
\end{equation}
(Such a function was used previously in \cite{Vol}.) Substitution of (\ref{phi}) into (\ref{EqD1}) gives the equation for $\varphi$
\begin{equation}
\partial _{i}\left [\varepsilon\partial _{i}  \varphi\right]
 =-4\pi \delta \left( {\bf r}-{\bf r}^{\prime }\right)\ .
\label{stat}
\end{equation}
This is the equation for the potential of a unit charge placed at point ${\bf r}^{\prime }$.

It is worth noting that one can derive these results by performing from the very beginning a gauge transformation under the $D$-function of the 4-vector-potential. Above we used the gauge \cite{DP}, where $D_{00}=D_{0i}=0$. However, in the problems where retardation effects are small, the ``Coulomb'' gauge, where  $D_{0i}=0$ and $\partial_i D_{ik}=0$, is more proper. In this gauge the main contribution gives $D_{00}$, which actually is equal to $-\varphi$ (compare  \S\   85, \cite{LP9}).

We can now express the interaction between an atom and body in the terms of the function $\varphi$:
\begin{equation}
V(l)=\frac{k_BT}{2}\alpha (0 )\left[
\partial _{i}\partial
_{i}^{\prime }\varphi\left( 0;{\bf r},{\bf r}^{\prime }\right) \right]
_{{\bf r}\rightarrow {\bf r}^{\prime }\rightarrow {\bf r}_{a}}\ . \label{Vgen1}
\end{equation}
This equation is valid for for any body. Therefore it describes all effects of the body size and shape. The function $\varphi $ must satisfy
usual electrostatic boundary conditions on the surface of the body. 
Then the boundary conditions for ${\cal D}_{ik}^{E}$ will be also satisfied. If an atom interacts with a body, which
occupies the half-space $z<0$, the boundary conditions are
\begin{equation}
\varphi _{z\rightarrow -0}=\varphi _{z\rightarrow +0},\varepsilon(0) \left(
\partial _{z}\varphi \right) _{z\rightarrow -0}=\left( \partial _{z}\varphi
\right) _{z\rightarrow +0}\ .
\end{equation}

The electrostatic problem is solved in Problem 1 to \S\ 7, \cite{LL8}. The solution is:
\begin{equation}
\varphi =\frac{1}{\left| {\bf r}-{\bf r}^{\prime }\right| }-\frac{%
\varepsilon -1 }{\varepsilon+1 }{\left[ \left( z+z^{\prime
}\right) ^{2}+\left( {\bf x}-{\bf x}^{\prime }\right) ^{2}\right] ^{-1/2}}
\label{phi1}
\end{equation}
where ${\bf r}=\{z,{\bf x}\}$.

The first term in (\ref{phi1}) does not depend on the presence of the dielectric and must be omitted. Differentiating the second term in order to calculate
${\cal D}_{lk}^{EL}\left( \zeta _{s};{\bf r},{\bf r}^{\prime }\right) $ and going to the limit ${\bf x\rightarrow x}^{\prime },$ $z\rightarrow z^{\prime
}\rightarrow l,$ we find after simple calculations
\begin{equation}
{\cal D}_{ii}^{EL}\left( 0;{\bf r,r}\right)=\hbar \partial_{l}\partial'_{l}\varphi\left( {\bf r},{\bf r}^{\prime }\right) =-\hbar
\frac{\varepsilon \left( 0 \right) -1}{\varepsilon
\left( 0 \right) +1}\frac{1}{2l^{3}}\ .  \label{DEf}
\end{equation}
 Equation (\ref{VLif}) follows immediately after substitution of (\ref{DEf}) into (\ref{Vgen1}).

Notice that the results which we discussed are valid only in the state of full thermodynamic equilibrium. In particular, (\ref{Vgen1}) is valid only if the temperature of the body is equal to the temperature of the black-body radiation that falls on the body from the $z>0$ half-space. If these two temperatures are different, the interaction energy decays as $l\rightarrow
\infty $ according to a slow $1/l^{2}$ law (see \cite{APS2}).

{\it Bad conductors.} The calculation of the contribution of free carriers in the longitudinal Green's function demands a microscopic approach. We will assume that the gas of carriers is not degenerate, i. e. obeys the Boltzmann statistics and, for simplicity, that the carriers have unit charges $\pm {\rm e}$ (${\rm e}$ is the elementary charge). Then perturbation of the density of, for example, the positive carriers produced by the static potential $\varphi$ is $\delta n^{(+)}=n^{(+)}\left (e^{-{\rm e}\varphi/k_{B}T}-1 \right) \approx -(n^{(+)}{\rm e}\varphi)/({k_{B}T)}$. Addition the corresponding charge density into (\ref{stat}) gives, for $z<0, z'>0$, the following equation for $\varphi$:
\begin{equation}
[\Delta  -\kappa^{2}]\varphi=0
\label{cond}
\end{equation}
where
\begin{equation}
\kappa^{2}=\frac{4\pi {\rm e}^2 n}{{\bar \varepsilon } k_BT}
\label{kappa}
\end{equation}
(compare \S\ 78, \cite{LL5}). Here  $n=n^{(-)}+n^{(+)}$ is the total density of current carriers  and ${\bar \varepsilon }$ is the ``bare'' dielectric constant, without contributions from the carriers.  Note that $\kappa=1/R_D$, where $R_D$ is the Debye radius. The Debye radius in good metals is of the order of interatomic distances. However, in ``bad'' conductors, where number of carriers is small, it can be comparable with the distance between the atom and surface. Equation (\ref{cond}) describes the screening of the electric field by the free carriers in the conductor. Indeed, according to this equation, the potential around a unit point charge in a uniform medium is $e^{-\kappa r}/{\bar \varepsilon }r$.  Notice that this equation can be interpreted in terms of spatial dispersion by introducing the longitudinal dielectric permittivity, which depends on the wave vector: $\varepsilon^L(k)={\bar \varepsilon }[1+1/(kR_D)^2]$. Accounting for the spatial dispersion is a difficult problem for restricted bodies. (See, for example \cite{Sernelius} and a review
\cite{Esquivel} and references therein. In these papers the effects of the spatial dispersion on interaction between conducting bodies are discussed.) It was easily solved in our static case due to the local connection between the electrostatic potential and the carrier density.

Thus the function $\varphi$ satisfies (\ref{stat}) at $z>0$ and (\ref{cond}) at $z<0$. On the boundary $z=0$ we have now the ``microscopic'' boundary conditions
\begin{equation}
\varphi _{z\rightarrow -0}=\varphi _{z\rightarrow +0},{\bar \varepsilon } \left(
\partial _{z}\varphi \right) _{z\rightarrow -0}=\left( \partial _{z}\varphi
\right) _{z\rightarrow +0}\ .
\label{bound1}
\end{equation}
To solve the equation, let us expand $\varphi({\mathbf x},z,z')$ into a Fourier integral with respect to ${\mathbf x}$:
\begin{equation}
\varphi_{\mathbf k}(z,z')=
\int\varphi({\mathbf x},z,z')e^{-i{\mathbf k}\cdot {\mathbf x}}d^2x\ .
\end{equation}

Let us consider first the $z>0$ domain. The Fourier transform of Eq.(\ref{stat}) is
\begin{equation}
[\partial^2_{z}  -k^{2}]\varphi_{\mathbf k}=-4 \pi \delta(z-z')\ .
\label{fk1}
\end{equation}
The presence of the $\delta$-function  means that $\varphi_{\mathbf k}$ at $z \to z'$ must have the singularity of the type $2\pi |z-z'|$. One easily finds the solution:
\begin{equation}
\varphi_{{\mathbf k}}=-2\pi\frac{e^{-k|z-z'|}}{k} + Ae^{-kz} \ .
\label{vac1}
\end{equation}
where $A(z')$ must be defined from the boundary conditions.

In the domain $z'>0, z<0$ the equation for $\varphi_{{\mathbf k}}$ can be obtained by the Fourier transform of (\ref{cond}):
\begin{equation}
[\partial^2_{z}  -q^{2}]\varphi_{\mathbf k}=0
\label{cond1}
\end{equation}
with
\begin{equation}
q=\sqrt{k^2+\kappa^2} \ .
\end{equation}
The solution is
$\varphi_{{\mathbf k}}= Be^{qz}  $.
Substitution of the boundary conditions (\ref{bound1}) gives the equations for $A$ and $B$:
\begin{equation}
\begin{split}
B=-2\pi e^{-kz'}/k+A  \\
{\bar \varepsilon} q B=-2\pi e^{-kz'}-kA \ .
\label{cond2}
\end{split}
\end{equation}
This gives
\begin{equation}
A(z')=\frac{2\pi}{k} \frac{{\bar \varepsilon} q-k}{{\bar \varepsilon} q+k}e^{-kz'} \ .
\end{equation}
We can omit now the ``free space'' first term in (\ref{vac1}) and write  the  function $\varphi_{{\mathbf k}}$ as
\begin{equation}
\varphi_{{\mathbf k}}(z,z')=\frac{2\pi}{k} \frac{{\bar \varepsilon} q-k}{{\bar \varepsilon} q+k}e^{-k(z+z')} \ .
\label{phikfin}
\end{equation}

Notice that $\varphi_{{\mathbf k}}$ depends on $z,z'$ only in the combination $z+z'$. This means that $\partial_{l}\partial'_{l}\varphi=[\partial_{z}^2-\partial_{\mathbf x}^2]\varphi$, or for the Fourier components $[\partial_{l}\partial'_{l}\varphi]_{\mathbf k}=-2k^2\varphi_{\mathbf k}$. We can put now $z=z'=l, {\mathbf x}={\mathbf x}'$.  Finally the potential of interaction  (\ref{Vgen1}) can be written as
\begin{equation}
V(l)=-
{k_BT}\alpha (0)\int_0^{\infty} \frac{{\bar \varepsilon} q-k}{{\bar \varepsilon} q+k}e^{-2kl}k^2dk \ .
\label{Vfin}
\end{equation}

If $\kappa \ll l^{-1}$, which corresponds to the case of a ``very bad'' conductor, $q \to k$ and we recover the result (\ref{VLif}) for an ideal dielectric, changing only $\varepsilon(0) \to {\bar \varepsilon}$. In the opposite limit $\kappa \gg l^{-1}$ (\ref{Vfin}) is reduced to the ``good'' metal result (\ref{VLifcond}). Equation (\ref{Vfin}) can be written as
\begin{equation}
V\left( l\right) =-\frac{k_BT}{4l^{3}}\alpha \left( 0\right)F_0(\xi) \ ,
\label{Vfin1}
\end{equation}
where $\xi =\kappa l=(l/R_D)$ and $F_0(\xi)$ is defined as
\begin{equation}
F_0(\xi)=
\frac{1}{2}
\int_0^{\infty} \frac{{\bar \varepsilon} \sqrt{4\xi^2+t^2}-t}{{\bar \varepsilon} \sqrt{4\xi^2+t^2}+t}e^{-t}t^2dt \ .
\label{F0}
\end{equation}

\begin{figure}[t]
\includegraphics[width=7cm,height=7cm]{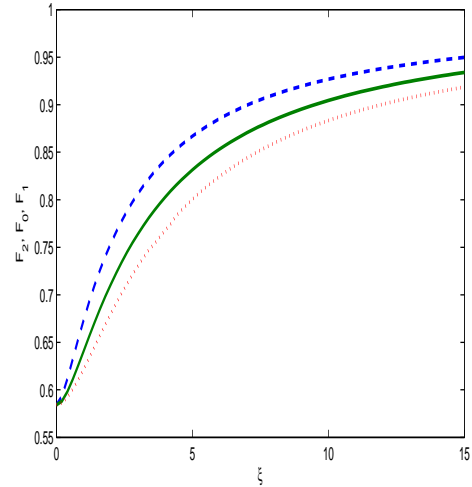}
\caption{Effect of the field penetration  in a silica sample on the atom-surface interaction.
Functions $F_0(\xi)$ - solid line,  $F_1(\xi)$ - dashed line and $F_2(\xi)$ - dotted line.}
\label{fig1}
\end{figure}

From an experimental point of view the more interesting quantities are the force acting an atom $-\partial_zV$ and its derivative. They can be presented as
\begin{equation}
-\partial_l V =-\frac{3k_BT}{4l^{4}}\alpha \left( 0\right)F_1(\xi) \ ,
\partial_l^2 V =-\frac{3k_BT}{l^{5}}\alpha \left( 0\right)F_2(\xi) \ .
\label{dV}
\end{equation}
One can easily find expressions for these functions by differentiating Eq. (\ref{phikfin}).
According to the definition these functions $F_i \to 1$  for $\xi \to \infty $ and $F_i \to ({\bar \varepsilon}-1)/({\bar \varepsilon}+1)$ for $\xi \to 0 $. Functions $F_i(\xi)$ are presented in Figure~\ref{fig1} for ${\bar \varepsilon}=3.81 $ (fused Silica).    One can notice the relatively slow convergence to the metal value at large $l$. 

Unfortunately the density of carriers in dielectrics depends very much on the technology of preparation of the samples and was not investigated systematically. For example, in \cite{Silica4}  values of $n$ in the interval between $10^{8}$ and $10^{15}$ cm$^{-3}$ were used for fused silica.  Then $R_D$ is in the interval $2.3 \times 10^{2} - 7.4\times 10^{-2}\mu\rm{ m}$.  Properties of the {\it vitreous silica} used in the experiments \cite{Harber,Obrecht07} are reviewed  in an article \cite{Vitr-Sil}. This medium is an ionic conductor, the main carriers are ions Na$^{+}$. The density of Na impurities is about $n_{Na}=3 \times 10^{15}$ cm$^{-3}$, however it is difficult to estimate the number of ions which are effective in mobility and screening.  It is not presently clear if the presence of  carriers has any influence on the interpretation of the results of the measurements \cite{Harber,Obrecht07}. (This question was discussed in \cite{Most1}.)  Notice that the relaxation time of the charge distribution is very long in bodies with small $\sigma$. This time can be calculated as $\tau_{c}={\bar \varepsilon}/(4 \pi \sigma)$. At room temperature the resistivity $\rho=1/\sigma \sim 10^{19} $ohm.cm$=1.1 \times 10^7 $s (see \cite{Fused}). This gives $\tau_{c} \sim 3.3 \times 10^{6} $s$=917$ hours. I believe, that at a such slow relaxation, the carriers mobility can hardly be important in any experiments.

In conclusion, a theory for the interaction between an atom and a conductor due to the thermal fluctuations is developed. The theory takes into account the  partial  penetration of static electric fluctuations into the conductor and is based on the Green's function technique in the presence of a spatial dispersion. A continuous crossover between an ideal dielectric and a good metal is investigated.

I thank M.~Antezza, Yu.~Barash,  D.~Dalvit, C.~Duran, S.~Ellingsen,  S.~Lamoreaux, L.~Pavesi, B.~Shklovskii, V.~Svetovoy, V.~Timofeev and particularly J.~Obrecht 
 for discussions  and G.~Klimchitskaya and V.~Mostepanenko for sending preprint \cite{Most1} before submitting.

\end{document}